\documentclass[titlepage,amsmath,amssymb]{article}
\usepackage{graphicx}% Include Figure files
\usepackage{dcolumn}% Align table columns on decimal point
\usepackage{bm}% bold math
\usepackage{latexsym}
\begin{document}
%%%%%%%%%%%%%%%%%%%%%%%%%%%%%%%%%%%%%
\title{Michaelis-Menten at 100 and allosterism at 50: driving molecular motors in a hailstorm with noisy ATPase engines and allosteric transmission} 
\author{Debashish Chowdhury \\ Department of Physics, Indian Institute of Technology,\\
        Kanpur 208016, India}
\date{\today}
\maketitle
%%%%%%%%%%%%%%%%%%%%%%%%%%%%%%%%%%%%%%%%%%%%%%%%%%%%%%%%%%%%%%%
\begin{abstract}
Cytoskeletal motor proteins move on filamentous tracks by converting input chemical energy that they derive by catalyzing the hydrolysis of ATP. The ATPase site is the analog of an engine and hydrolysis of ATP is the analog of burning of chemical fuel. Moreover, the functional role of a segment of the motor is analogous to that of the  transmission system of an automobile that consists of shaft, gear, clutch, etc. 
The operation of the engine is intrinsically ``noisy'' and the motor faces a molecular ``hailstorm'' in the aqueous medium. In this commemorative article, we celebrate the centenary of Michaelis and Menten's landmark paper of 1913 and the golden jubilee of Monod et al.'s classic paper of 1963 by highlighting their relevance in explaining the operational mechanisms of the engine and the transmission system, respectively, of  cytoskeletal motors. 

\end{abstract}
%%%%%%%%%%%%%%%%%%%%%%%%%%%%%%%%%%%%%%%%%%%%%%%%%%%%%%%%%%%%%%%

%%%%%%%%%%%%%%%%%%%%%%%%%%%%%%%%%%%%%%%%%%%%%%%%%%%%
\section{Introduction}
%%%%%%%%%%%%%%%%%%%%%%%%%%%%%%%%%%%%%%%%%%%%%%%%%%%%

In 1913 Leonor Michaelis and Maud Menten published a paper \cite{michaelis1913} (see ref.\cite{johnson11} for english translation) in which they derived an analytical expression (from now onwards referred to as the MM equation) for the rate of an extremely simple model of enzymatic reactions \cite{cornishbowden}. 
Even after enriching several overlapping scientific disciplines for 100 years, MM kinetics still continues to raise new exciting fundamental questions on the conditions for its own validity as well as on its applicability to hitherto unexplored situations. One of the systems where such issues are being debated is molecular motors \cite{keller00,howard01,mogilner02,block07,fisher07,chowdhury13a,chowdhury13b}.

These chemo-mechanical machines are either single proteins or macromolecular complexes; their stepping on the respective filamentous tracks are random, albeit biased, and these run on ``noisy engines''.  For the sake of concreteness in this article we'll consider mostly cytoskeletal motors whose engines are  ATPases   (more precisely, the ATP-hydrolyzing  sites in the head domains of the motors). These motors drive many processes that are involved in sub cellular and cellular motility as well as contractility, including mitosis and cytokinesis \cite{chowdhury13a,chowdhury13b,pollard03a}. The ``transmission system'' of the motor amplifies the sub-nanometer movements, powered by ATP hydrolysis, in the engine to several nanometer long step size along its track. Moreover, the ATPase cycle of the engine also regulates the track-binding affinity of the motor whose track-binding site is located a few nano-meters away from the engine. Long-distance intra-machine communications required for track-binding regulation and for the transmission system operation are based on allosteric mechanisms. The basic principle of allostery was laid down 50 years ago in a classic paper by Monod, Changeux and Jacob \cite{monod63}.

The aim of this commemorative note is to celebrate the centenary and golden jubilee of the pathbreaking papers of Michaelis and Menten \cite{michaelis1913} and Monod et al.\cite{monod63}, respectively,  by (a) highlighting some important features of the fluctuating ATPase activity of the engine and the allosteric mechanisms of the transmission system of a molecular motor, and (b) exploring the dependence of the velocity of the motor on the rate of ``fuel consumption'' by its engine. We also point out some interesting formal analogies between the enzymatic turnover and the mechanical stepping of a molecular motor. We emphasize only the main conceptual developments over the last two decades, but make no attempt to cover all aspects of molecular motors; interested readers are referred to two very recent review articles \cite{chowdhury13a,chowdhury13b}.

%%%%%%%%%%%%%%%%%%%%%%%%%%%%%%%%%%%%%%%%%%%%%%%%%%%%%%%%
\section{Key mechanical properties of molecular motors and core concepts in their operational mechanisms} 
%%%%%%%%%%%%%%%%%%%%%%%%%%%%%%%%%%%%%%%%%%%%%%%%%%%%%%%%%

The tracks for molecular motors are polar, that is the two ends of these filaments are not identical. On a given track, the members of the same family of motors tend to move in the same direction. The more processive a motor is the longer is the distance it covers in a single run in between its attachment to the track and the next complete detachment from it. Thus, each family of motors, that move on a given filamentous track, is characterized by a specific {\it directionality} and {\it processivity}. The fraction of the cycle time that each head remains attached to the track, on the average, is called its {\it duty ratio}.

An external force on a motor against the direction of its natural stepping is called a {\it load force}. The average velocity of a motor decreases with increasing magnitude of the load force and eventually vanishes at a value that is called the {\it stall force}. The {\it force-velocity relation} is one of the most fundamental characteristics of a molecular motor. Since the stepping of molecular motors is noisy, two different motors with identical mean velocities may exhibit widely different variances. Therefore, the stochastic stepping kinetics of a motor is characterized in more detail in terms of the {\it distributions of the times of its dwell} at the successive spatial positions on its track.

Cytoskeletal motor proteins, which are listed in table \ref{table-porters}, function as intracellular ``porters'' \cite{leibler93}. These motors carry intracellular cargoes (e.g., vesicles, organelles, etc.) over long distances by ``walking'' along their respective tracks hydrolyzing ATP  which is the most widely used ``fuel'' for molecular machines. In contrast, the motors listed in the table \ref{table-rowslide} slide one cytoskeletal polar filament with respect to another. The sliding occurs when a slider motor, that crosslinks the two filaments, tends to walk on both simultaneously hydrolyzing ATP. Some {\it sliders} work in groups and each detaches from the filament after every single stroke. These are often referred to as {\it rowers} because of the obvious analogy with rowing where the oars remain in contact with water for a very brief period in each stroke \cite{leibler93}. Sliders and rowers drive {\it contractility} at cellular and subcellular levels. It is worth pointing out that occasionally a molecular motor can undergo a mechanical transition that requires more work than the free energy supplied by the hydrolysis of one molecule of ATP. The molecular motor accomplishes such a rare feat, that no macroscopic motor can perform, because of fluctuations arising from its interactions with the thermal reservoir.

Polynucleotides (DNA and RNA) and polypeptides (proteins) are linear polymers. The sequence of the monomeric subunits of each of these polymers is dictated by that of the corresponding template. The sequence of the subunits of a template directs the correct sequence of the successive monomers that are to be selected, and added step-by-step, by the machine thereby elongating the product polymer. Specific machine to be used for a template-directed polymerization depends on the nature of the template and product polymers; these are listed in the table \ref{table-polymerase}. The template also serves as the track for the polymerizer machine and its movement along the track is powered by input chemical energy. Therefore, these machines are also regarded as motors\cite{cozzarelli06a,dulin13}.

MT and F-actin serve not only as tracks for cytoskeletal motors, but can also generate pushing and pulling forces during their polymerization and depolymerization, respectively. For example, a polymerizing MT can exert a pushing force against a membrane thereby mimicing a nano-piston \cite{chowdhury13b}. Similarly, during its depolymerization a  MT can pull a molecular ring by inserting its hook-like outwardly curled depolymerizing tip into the ring \cite{chowdhury13b}. In addition to the motors listed above, cell uses also several other types of motors for various specialized functions; a more exhaustive list is available in ref.\cite{chowdhury13a}.

%%%%%%%%%%%%%%%%%%%%%%%%%%%%%%%%%%%%%%%%%%%%%%%%%%%%%%
\section{Average rate of a MM reaction: implications for the average velocity of a motor}
%%%%%%%%%%%%%%%%%%%%%%%%%%%%%%%%%%%%%%%%%%%%%%%%%%%%%%

The MM reaction is normally represented as 
\begin{eqnarray} 
{\cal E} + {\cal S} \mathop{{\rightleftharpoons}}^{k_1}_{k_{-1}} {\cal ES} \mathop{{\rightarrow}}^{k_2} {\cal E} + {\cal P}  
\label{eq-MM}
\end{eqnarray}
where ${\cal E}, {\cal S}, {\cal P}$ denote the enzyme, substrate and product, respectively. 
The most commonly used expression for $V$, the rate of the reaction (\ref{eq-MM}) in bulk, was derived, 
more than a decade after Michaelis and Menten's original derivation, under a less restrictive approximation. In that work \cite{briggs1925}, Briggs and Haldane assumed that the concentration of the intermediate complex $[ES]$ remains steady (or, stationary), i.e., independent of time, during the progress of the reaction in bulk. One of the standard forms in which the corresponding $V$ is expressed is 

\begin{equation}
V = \frac{V_{max} [S]}{K_{m} + [S]}
\label{eq-MMrate}
\end{equation}
which, in addition to the substrate concentration $[S]$, involves two important parameters $K_M$ and $V_{max}$. The parameter  $V_{max} = k_2 [{\cal E}]_0$ is proportional to the total amount of 
enzyme $[{\cal E}]_{0}$ and it can be interpreted as the maximum rate of the reaction attainable in the limit $[S] \to \infty$. The other parameter 
\begin{equation}
K_M = \frac{k_{-1}+k_2}{k_1},  
\label{eq-michconstant}
\end{equation}
called the {\it Michaelis constant}, can be interpreted as the particular concentration of substrate for which the rate of the reaction is exactly half of its maximum possible value. 

The importance of Michaelis and Menten's classic work \cite{michaelis1913} can be appreciated in its historical context  by revisiting the current state of that research field in 1913 (see, for example, the 
review by  Barendrecht \cite{barendrecht1913}). Michaelis and Menten's achievement also demonstrated how, inspired by the correct insight, even a minimal theoretical model and an extremely simplifying assumption can still account for empirical data obtained for varieties of real systems under wide range of conditions \cite{gunawardena12}. Whether or not real enzymes follow the equation 
(\ref{eq-MMrate}) has been scrutinised in the last 100 years by analysing the biochemical data for an enormously large number of enzymes (see refs. \cite{hill77,schnell03}). The history of the kinetics of enzymatic reactions over the last 100 years \cite{gutfreund76,laidler97,baici06}  is testimony to the 
success of MM reaction and its various extensions. 

\begin{figure}[htbp]
\begin{center}
\includegraphics[width=0.95\columnwidth]{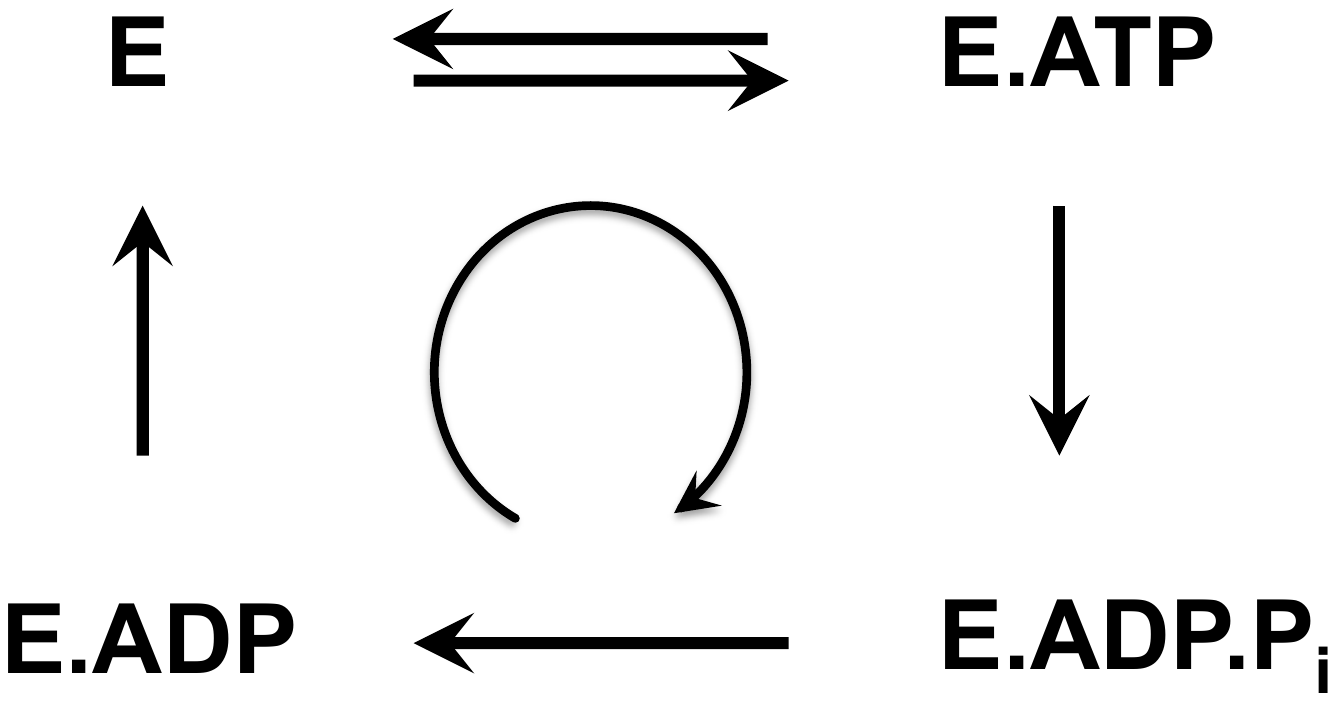}
\end{center}
\caption{The main pathway of ATP hydrolysis by an ATPase (ATP-hydrolyzing enzyme). 
See the text for explanation.
}
\label{fig-atphydro}
\end{figure}

Since a cytoskeletal motor  runs on an ATPase engine, one would like to know the dependence of the average velocity $v$ of the motor on the ATP concentration in the surrounding medium. 
the enzymatic cycle can be expressed symbolically as (see fig.\ref{fig-atphydro}) 
\begin{equation}
E + ATP \rightleftharpoons E.ATP \rightarrow E.ADP.P_{i} \rightarrow E.ADP \rightarrow E  
\label{eq-ATPcycle}
\end{equation}
where $E.ATP$ is a complex of the enzyme $E$ and an ATP molecule whereas $E.ADP.P_{i}$ and $E.ADP$ are macromolecular complexes of the enzyme with the products of hydrolysis. In this case, in each cycle one molecule of ATP is hydrolyzed to produce one molecule of ADP and one inorganic phosphate molecule.

If no futile reaction takes place, i.e., every round of chemical reaction drives mechanical stepping, the chemical and mechanical processes are said to be tightly coupled. For a tightly coupled molecular motor, $v$ should be given by $v = V L$ where $L$ is the step size of the motor and $V$ is the average rate of the ATPase reaction given by equation (\ref{eq-MMrate}) with $[S]=[ATP]$. Therefore, $v$ should saturate with increasing concentration of ATP. 
In reality,  for many motors the chemical and mechanical processes are coupled loosely \cite{knight99,oosawa00,mehta01,clancy11}.
In case of loosely coupled motors, it has been observed that $v=\kappa V L$ where $\kappa < 1$ is the strength of the mechano-chemical coupling. What is even more surprising is that even when the natural forward movement of the motor is opposed by a load force $F$, a MM-like form 
\begin{equation}
v(F) = \kappa(F) L \frac{V_{max}(F) [ATP]}{K_{m}(F) + [ATP]}
\label{eq-block} 
\end{equation}
seems to capture the dependence of $v$ on the ATP concentration. This is, perhaps, not the appropriate forum for  detailed discussion \cite{howard01} to find out which, if any, of the parameters $\kappa(F), V_{max}(F)$ and $K_{m}(F)$ are independent of the load force $F$. 
Nevertheless, if an individual rate constant for a specific mechano-chemical step, indeed, varies with the load force $F$ what is the form of that dependence? In the literature on molecular motors, the force-dependence of the individual rate constants associated with the forward and reverse transitions, at a given ATP concentration, are assumed to have the form \cite{fisher07,chowdhury13a}
\begin{equation}
k_{f}(F) = k_{f}(0) e^{-\theta  F(\Delta x)/(k_BT)}
\label{eq-fdepf}
\end{equation}
and
\begin{equation}
k_{r}(f) = k_{r}(0) e^{(1-\theta)  F(\Delta x)/(k_BT)},
\label{eq-fdepr}
\end{equation}
respectively, where $k_B$ is the Boltzmann constant, $T$ is the absolute temperature and $\theta$ is a fraction of the step size $\Delta x$. Note that $\theta$ determines how the work $F \Delta x$ performed by the external load is shared by the forward and reverse transitions \cite{fisher07}.

\begin{figure}[htbp]
\begin{center}
\includegraphics[width=0.95\columnwidth]{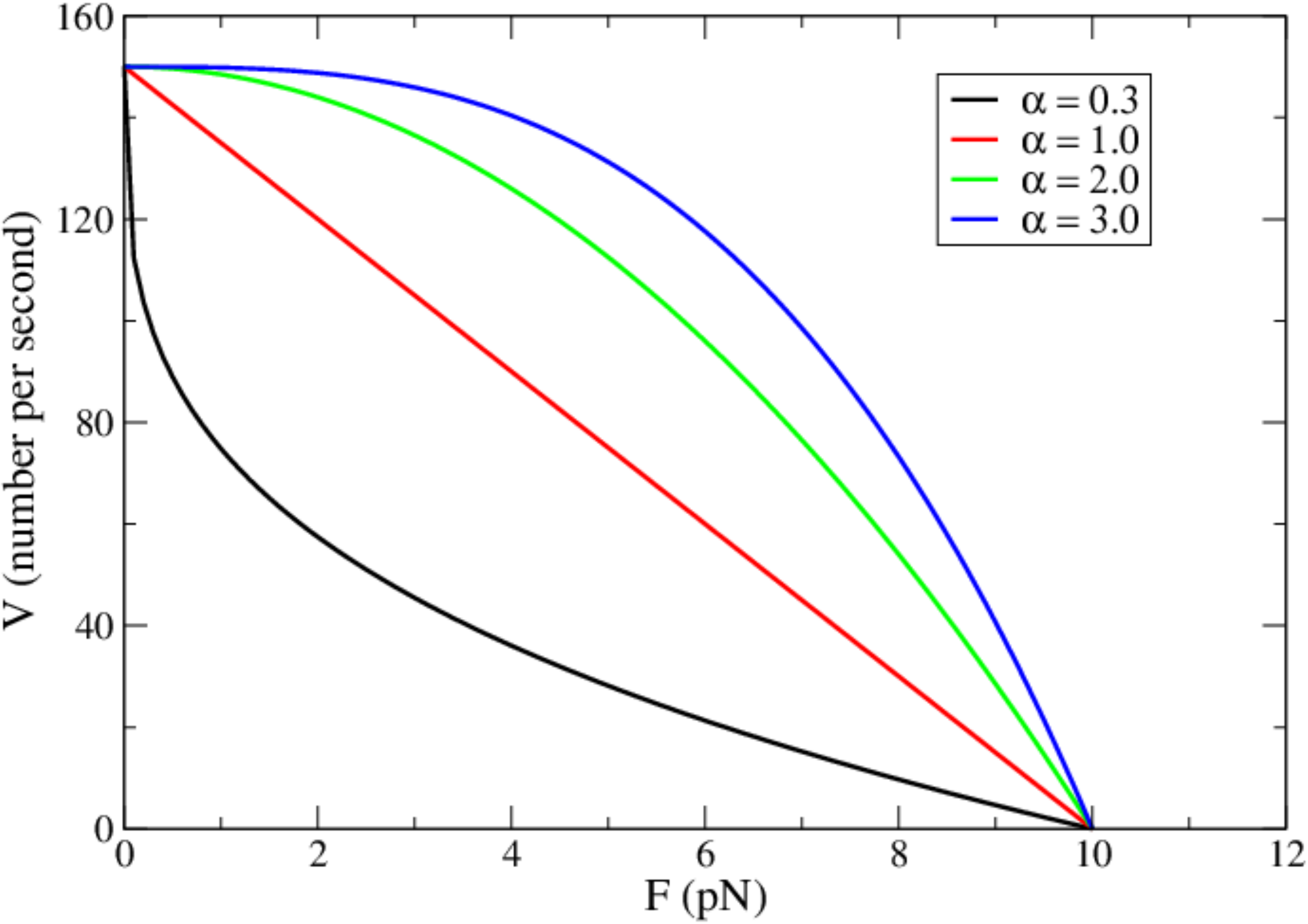}
\end{center}
\caption{Typical force-velocity relations of a molecular motor obtained by graphically plotting (\ref{eq-fvalpha}) for $v(0)=150$s$^{-1}$, $F_{s}=10$ pN and a four different values of the parameter $\alpha$. The unit ``number per second'' refers to the number of subunits of its track.
}
\label{fig-forcevel}
\end{figure}

The average velocity of a motor is, in general, a combination of the individual rate constants and, therefore, expected to depend on the load-force through the $F$-dependence of the rate constants. It has been established experimentally that, at a given concentration of ATP, $v(F)$ decreases with increasing magnitude of the load force $F$, and eventually vanishes at the stall force $F_{s}$. Both the load-free velocity $v(0)$ and the stall force $F_{s}$ depend on the details of the mechano-chemical kinetics of the motor. The experimentally observed force-velocity relation is often expressed in the simple form \cite{kunwar10}
\begin{equation}
v(F) = v(0) [1-(F/F_{s})^{\alpha}] 
\label{eq-fvalpha}
\end{equation}
in terms of the three parameters $v(0), F_{s}$ and $\alpha$. The force-velocity relation is linear in the special case $\alpha = 1$ whereas it is  {\it sub-linear} for all $\alpha < 1$ and {\it super-linear} for $\alpha > 1$. Note that the parameter $\alpha$ determines the curvature (i.e., convex-up or concave-up shape) of the force-velocity relation (see Fig.\ref{fig-forcevel}). Motors with superlinear force-velocity relation are strong in the sense that their velocity is practically unaffected by the load force unless the magnitude of the force is sufficiently large. On the other hand, if the force-velocity relation of a motor is sub-linear, even a small load force drastically reduces its velocity thereby exposing its weakness. 

It is worth pointing out that load force can affect the mechano-chemistry of the motor in many different ways. For example, a load force, even if weaker than the stall force, can increase the frequencies of back-stepping. It can also tilt the free energy landscape to such an extent that the motor steps backward while binding with ADP and inorganic phosphate, and it subsequently releases the reformed ATP.

What happens to the motor if the strength of the load force opposing it exceeds $F_{s}$? In principle, one can envisage three possible scenarios: (i) the motor may detach from the track, (ii) the motor may start moving backward hydrolysing one molecule of ATP in each backward step, or (iii) the motor start moving backward without hydrolyzing ATP as it gets pushed by the load force \cite{cross06}. 
Recal that $v(F)$ is the average velocity of the motor. Because of the stochastic nature of its mechano-chemistry, the three possible scenarios painted here can occur, with increasing frequency as the load forces increases, even if it remains weaker than $F_{s}$. The actual outcome depends on the family to which the motor belongs.

%%%%%%%%%%%%%%%%%%%%%%%%%%%%%%%%%%%%%%%%%%%%%%%%%%%%%%%%
\section{Distribution of turnover times of an enzyme: implications for the noisy ATPase engine of a motor} 
%%%%%%%%%%%%%%%%%%%%%%%%%%%%%%%%%%%%%%%%%%%%%%%%%%%%%%%%%

\begin{figure}[htbp]
\begin{center}
\includegraphics[width=0.95\columnwidth]{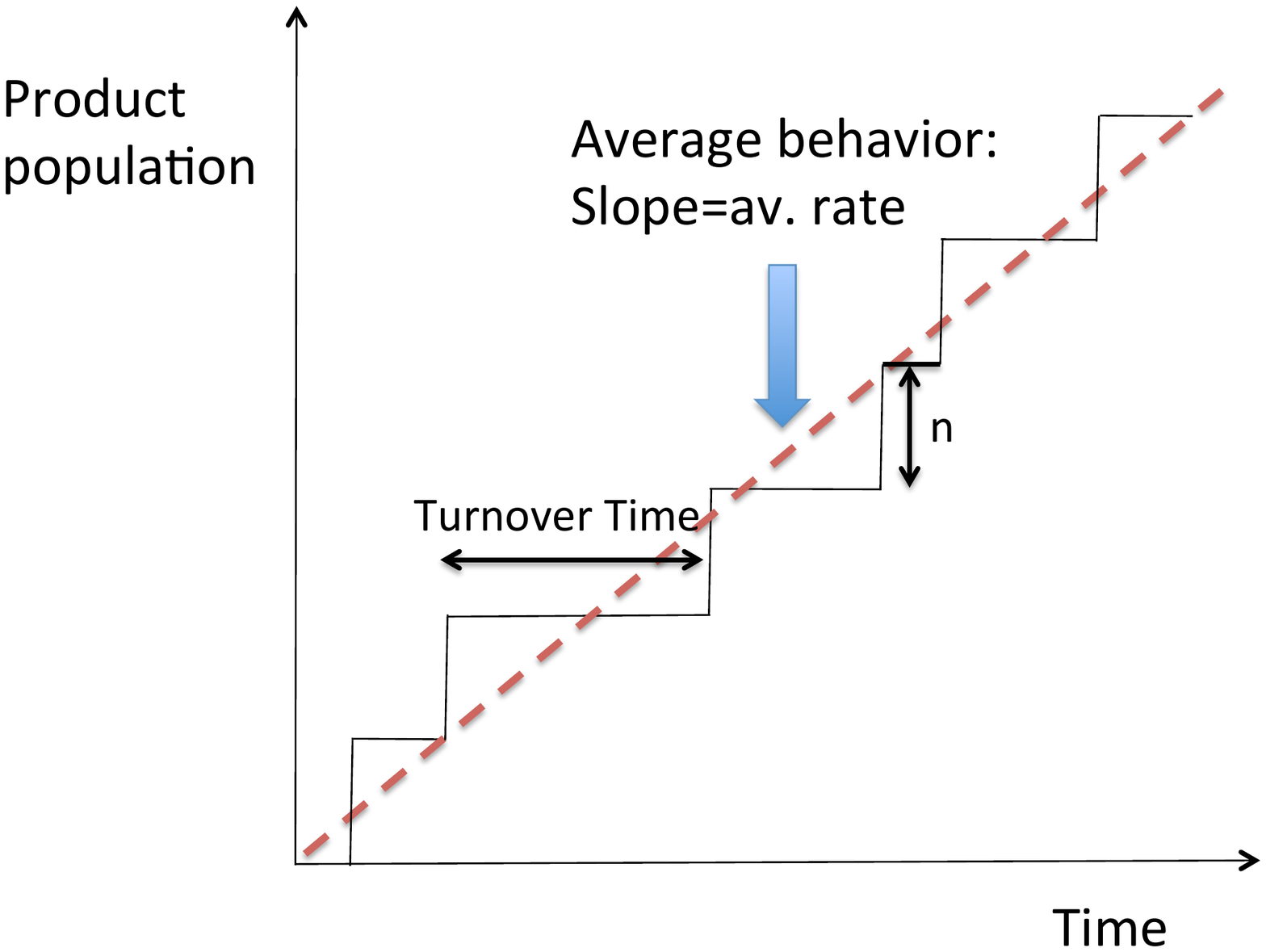}
\end{center}
\caption{A hypothetical noise-free trace of the population of the product molecules of an enzymatic reaction during multiple turnovers catalyzed by a single enzyme. The slope of the dashed line indicates the average rate of increase of the reaction product with the passage of time.
}
\label{fig-enzymeTIME}
\end{figure}

Next, we consider multiple turnovers by a single-enzyme. Obviously, there are nonzero waiting periods between the successive turnovers of the enzyme \cite{min05,kou05,min06}. The time interval $\tau$ in between the successive turnovers of an enzyme is called the turnover time \cite{kou08,yang11}. Suppose in each round of the reaction catalyzed by the enzyme $n$ product molecules are produced. Thus, the population of the product molecules increases by $n$ in each discrete jump (see Fig.\ref{fig-enzymeTIME}). In the specific case of the ATPase engine of a molecular motor, the time interval in between the arrival of the successive ATP molecules and their binding to the engine is random. Moreover, the actual time taken for conversion of the ATP molecule to the product(s) of hydrolysis also fluctuates from one round of the reaction to the next.  Furthermore, the conformation of a motor, like that of any other enzyme, itself fluctuates (the so-called dynamic disorder) giving rise to further sources of noise in the turnover time (see Fig.\ref{fig-enzymeTIME}). In general, the distribution of the turnover times characterizes the stochastic kinetics of the enzymatic reaction; the mean turnover time $<\tau>$ happens to be just the first moment of this distribution. 

In the last decade sophisticated techniques of single-molecule enzymology has made it possible to monitor the enzymatic turnover of a single enzyme \cite{xie10,moffitt13}. For technical reasons, turnover statistics has been studied experimentally using some non-motor enzymes although the conclusions drawn from these experiment are believed to be valid also for the ATPase activities of the engines of the motors. Interestingly, barring a few exceptional situations, the average rate $1/<\tau>$ of the enzymatic reactions obey the MM equation (\ref{eq-MMrate}) \cite{kou05,min06,kou08,yang11}. Thus, in spite of the fluctuations, the average rate of fuel burning by the noisy ATPase engine of a cytoskeletal motor is still given by the MM equation (\ref{eq-MMrate}).

%%%%%%%%%%%%%%%%%%%%%%%%%%%%%%%%%%%%%%%%%%%%%%%%%%%%%%%%
\section{Analogy between turnover time of enzyme and dwell time of motor: implications for the distributions} 
%%%%%%%%%%%%%%%%%%%%%%%%%%%%%%%%%%%%%%%%%%%%%%%%%%%%%%%%%

\begin{figure}[htbp]
\begin{center}
\includegraphics[width=0.95\columnwidth]{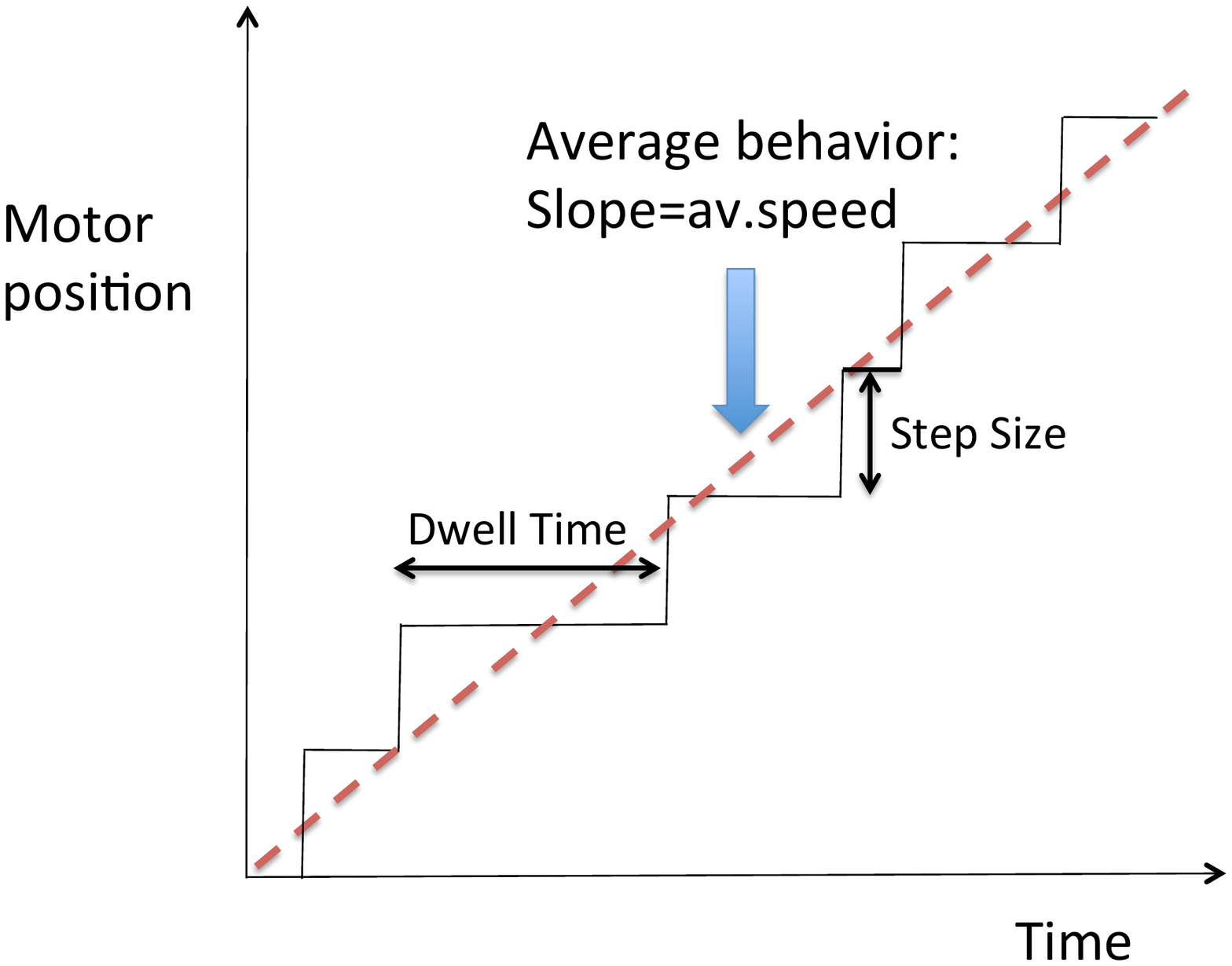}
\end{center}
\caption{A hypothetical noise-free trace of the positions of a molecular motor with the progress of time. The length of the vertical segments are the step sizes while that of the horizontal segments are the dwell times. No backward step is observed on this trace. The slope of the dashed line indicates the average velocity of the motor.
}
\label{fig-motorTIME}
\end{figure}

Next we discuss a formal analogy between the turnover times of the engine and the times of dwell of the motor at its successive positions on the track.  Single-motor experiments have demonstrated that the movement of each motor consists of an alternating sequence of pause and translocation. In other words, a motor dwells at each discrete position on its track for a duration in between its arrival at that position and the next departure from it (see Fig.\ref{fig-motorTIME}). The nature of the walk of a given motor is characterised by the distribution of the dwell times. In fact,  both these distributions also have a practical utility; the quantity 
\begin{equation}
n_{min} = <\tau>^{2}/(<\tau^2>-<\tau>^2)
\label{eq-nmin}
\end{equation} 
provides a lower bound on the number of kinetic states \cite{moffitt10b,moffitt13}. 
The similarities between the trace of successive turnovers in fig.\ref{fig-enzymeTIME} and the trace of successive steps of the motor in fig.\ref{fig-motorTIME} indicates the possibility of using the same mathematical formalism for extracting their statistical properties.

Next, we explicitly mention the nature of the dwell time distribution for motors with simple mechano-chemical kinetics and point out some of its key features.
For the operation of a molecular motor driven by an ATPase engine, the hydrolysis of ATP generates a mechanical force. However, this force may be generated at different stages of the enzymatic reaction for different motors. Assuming that the force is generated in only one single step of the enzymatic cycle, four distinct possible stages of force generation are listed below where the subscripts $j$ and $j+1$ label the two successive positions of the motor $M$ on its track \cite{keller00}.

\noindent Mechano-chemical binding:
$$(M + ATP)_{j} \rightleftharpoons (M.ATP)_{j+1} \rightleftharpoons (M.ADP.P_{i})_{j+1} \rightleftharpoons (M.ADP)_{j+1} \rightleftharpoons (M)_{j+1}$$
\noindent Mechano-chemical reaction: 
$$(M + ATP)_{j} \rightleftharpoons (M.ATP)_{j} \rightleftharpoons (M.ADP.P_{i})_{j+1} \rightleftharpoons (M.ADP)_{j+1} \rightleftharpoons (M)_{j+1}$$
\noindent Mechano-chemical release: 
$$(M + ATP)_{j} \rightleftharpoons (M.ATP)_{j} \rightleftharpoons (M.ADP.P_{i})_{j} \rightleftharpoons (M.ADP)_{j+1} \rightleftharpoons (M)_{j+1}$$
\noindent Mechano-chemical trigger: 
$$(M + ATP)_{j} \rightleftharpoons (M.ATP)_{j} \rightleftharpoons (M.ADP.P_{i})_{j} \rightleftharpoons (M.ADP)_{j} \rightleftharpoons (M)_{j+1}$$

\begin{figure}[htbp]
\begin{center}
\includegraphics[width=0.95\columnwidth]{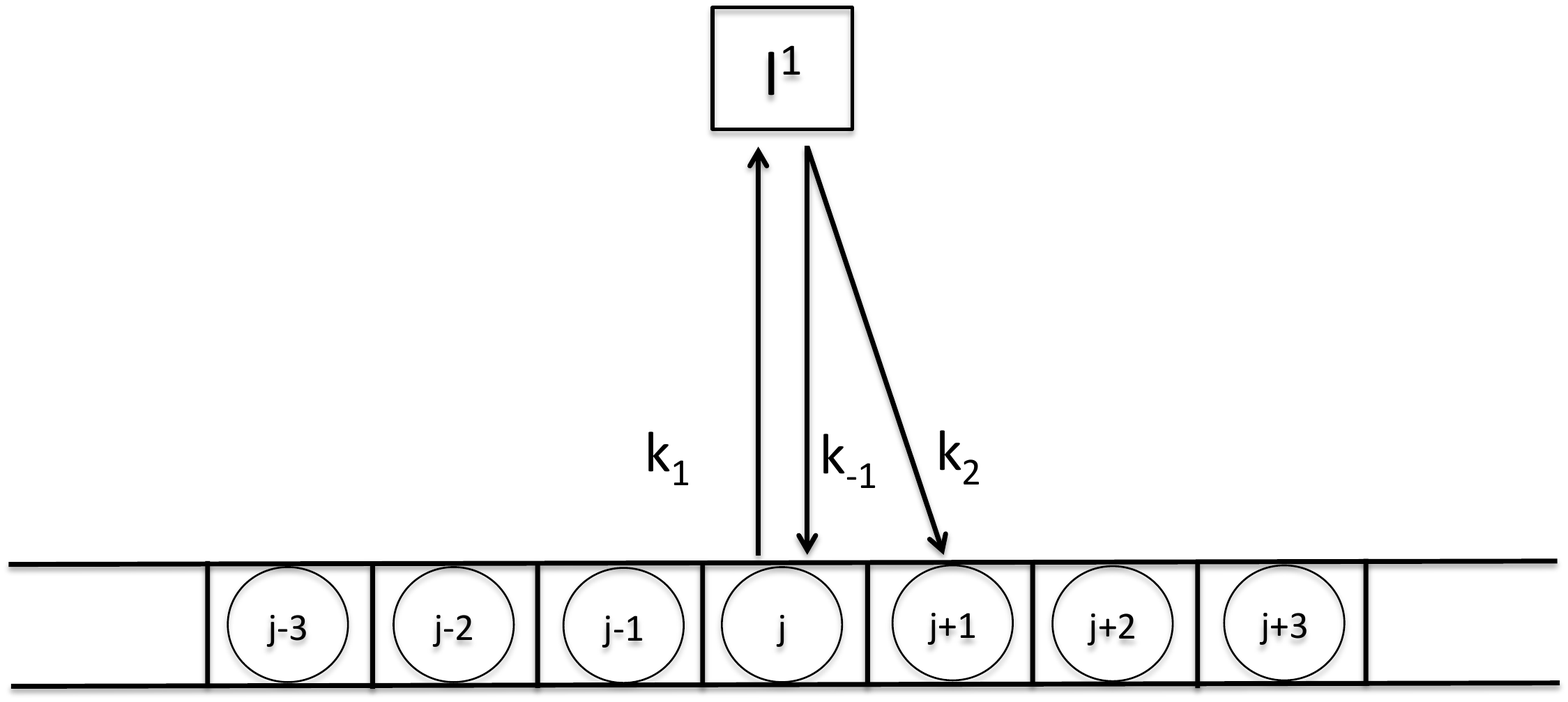}
\end{center}
\caption{A schematic representation of a simple mechano-chemical kinetics of a motor that walks on a linear track. The discrete allowed positions of the motor on the track are labelled by an integer j.  The motor can move only forward by one step; backward steps are not allowed. $I_{1}$ is an intermediate 
``chemical'' (or ``internal'') state. The vertical arrows indicate purely ``chemical'' transitions in which the mechanical position of the motor does not change whereas the transition represented by the tilted arrow is a mechano-chemical transition in which both the chemical state and mechanical position of the motor change simultaneously.
}
\label{fig-mottrack}
\end{figure}

The transition from position $j$ to position $j+1$ is the force-generating step of the cycle. Each of these four schemes is based on the assumption of tight mechano-chemical coupling although these can be easily extended to model loose-coupling motors. 
In all the four cases the average velocity of the motor is given by the same expression which, however, differs from the MM equation (\ref{eq-MMrate}); the only difference between the four cases is that different rate constants are force-dependent in the different models \cite{keller00}.  Moreover, under appropriate circumstances, by short-circuiting relatively fast transitions (i.e., by combining more than one step into a single effective step) the number of intermediate complexes can be reduced to only one. 
Furthermore, under some conditions one or more of the transitions can also become practically irreversible. For example, if the products of ATP hydrolysis are removed immediately after their formation, their rebinding with the motor can be ruled out. The typical example shown below
$$(M + ATP)_{j} \rightleftharpoons I_{j}^{1} \rightarrow (M)_{j+1},$$
where $I_{j}^{1}$ is the intermediate complex and the last transition has been assumed to be irreversible, 
mimics exactly the MM scheme (\ref{eq-MM}). This scheme has been depicted slightly differently in  Fig.\ref{fig-mottrack}, where the discrete allowed positions correspond to the motor-binding sites on the track. 

For the generic kinetics shown in Fig.\ref{fig-mottrack}, the probability density of the dwell times is given by 
\begin{equation}
P(t) = \biggl(\frac{k_1 k_2 }{2 B}\biggr)\biggl\{e^{-k_{-} t} - e^{-k_{+} t}\biggr\}
\label{eq-ftsingMM}
\end{equation}
where
\begin{equation}
k_{\pm} = A \pm B
\end{equation}
with 
\begin{eqnarray}
A &=& \frac{(k_{1}+k_{-1}+k_{2})}{2}    \nonumber \\ 
B &=& \biggl(\frac{(k_{1}+k_{-1}+k_{2})^{2}}{4}-k_{1} k_{2}\biggr)^{1/2}.
\label{eq-singMMB}
\end{eqnarray}
Note that the distribution $P(t)$ for the two-step process shown in Fig.\ref{fig-mottrack} is a sum of two exponentials where the rates $k_{\pm}$ in these two exponentials are linear combinations of the rate constants $k_{1}, k_{-1}, k_{2}$. In the special limiting case $k_{-1}=0$, the expression 
(\ref{eq-ftsingMM}) simples to 
\begin{equation}
P(t) = \biggl(\frac{k_1 k_2 }{k_1-k_2}\biggr)\biggl\{e^{-k_2t} - e^{-k_1t}\biggr\}
\label{eq-irrevft}
\end{equation}
where the two rates in the two exponentials are the two non-zero rate constants $k_1, k_{-1}$. 
Moreover, in this special limit the mean dwell time becomes $k_{1}^{-1}+k_{2}^{-1}$, i.e., sum 
of the average times spent sequentially in the two states.

Even among the molecular motors that never step backward and do not have branched 
pathways, very few have mechano-chemical kinetics as simple as that shown in 
Fig.\ref{fig-mottrack}. One such example is the ribosome, one of the largest and most 
complex molecular motors, that moves on a messenger RNA (mRNA) track while synthesising a 
polypeptide using its mRNA track as a template. Its mechano-chemical cycle, during the stage 
of elongation of the polypeptide, has a few more intermediate states after $I_{1}$.
However, in spite of the existence of more than one intermediate states, the average 
rate of elongation of the polypeptide (which is identical to the average velocity of the 
ribosome on its track) is governed by a MM-like equation \cite{garai09a} when kinetic 
proofreading and translational error are not captured explicitly by the model. This is not 
surprising because the average rate of a MM reaction with more than one intermediate 
complex is also known to obey a MM-like equation.  However, what is non-trivial is 
that even after inclusion of the pathways (a) for tRNA rejection by kinetic proofreading 
and (b) for incorporation of wrong amino acid into the growing polypeptide, the 
mean rate of elongation of the polypeptide still satisfies a MM-like equation 
\cite{sharma11a}.
But, there are few kinetic models of molecular motors, including one that we 
\cite{garai11} developed for the single-headed kinesin KIF1A, whose average velocity 
deviate from the MM-like behaviour. We'll not discuss the nature and plausible causes 
of these deviations here because the details are available in ref.\cite{moffitt13}. 

During protein synthesis {\it in-vivo} usually a large number of ribosomes simultaneously 
move on the same mRNA track, each polymerizing one copy of the same protein whose 
amino acid sequence is directed by the mRNA that serves also as a template. This 
phenomenon is often referred to as the ribosome traffic \cite{chowdhury13a} because 
of its superficial similarity with vehicular traffic on highways. A ribosome poised for 
forward translocation to the next codon has to wait at its present location, if the target 
codon is already covered by another ribosome in front of it. The following ribosome 
can move forward only after its target codon is vacated by the leading codon. Thus, 
increasing crowding lengthens the dwell time of a ribosome at a codon. The effect of 
the crowding on the overall distribution of the dwell times of the ribosomes has been 
calculated analyticaly \cite{sharma11a}.

%%%%%%%%%%%%%%%%%%%%%%%%%%%%%%%%%%%%%%%%%%%%%%%%%%%%%%
\section{Forward, backward and reverse processes: conditional dwell times}
%%%%%%%%%%%%%%%%%%%%%%%%%%%%%%%%%%%%%%%%%%%%%%%%%%%%%%

\begin{figure}[htbp]
\begin{center}
\includegraphics[width=0.95\columnwidth]{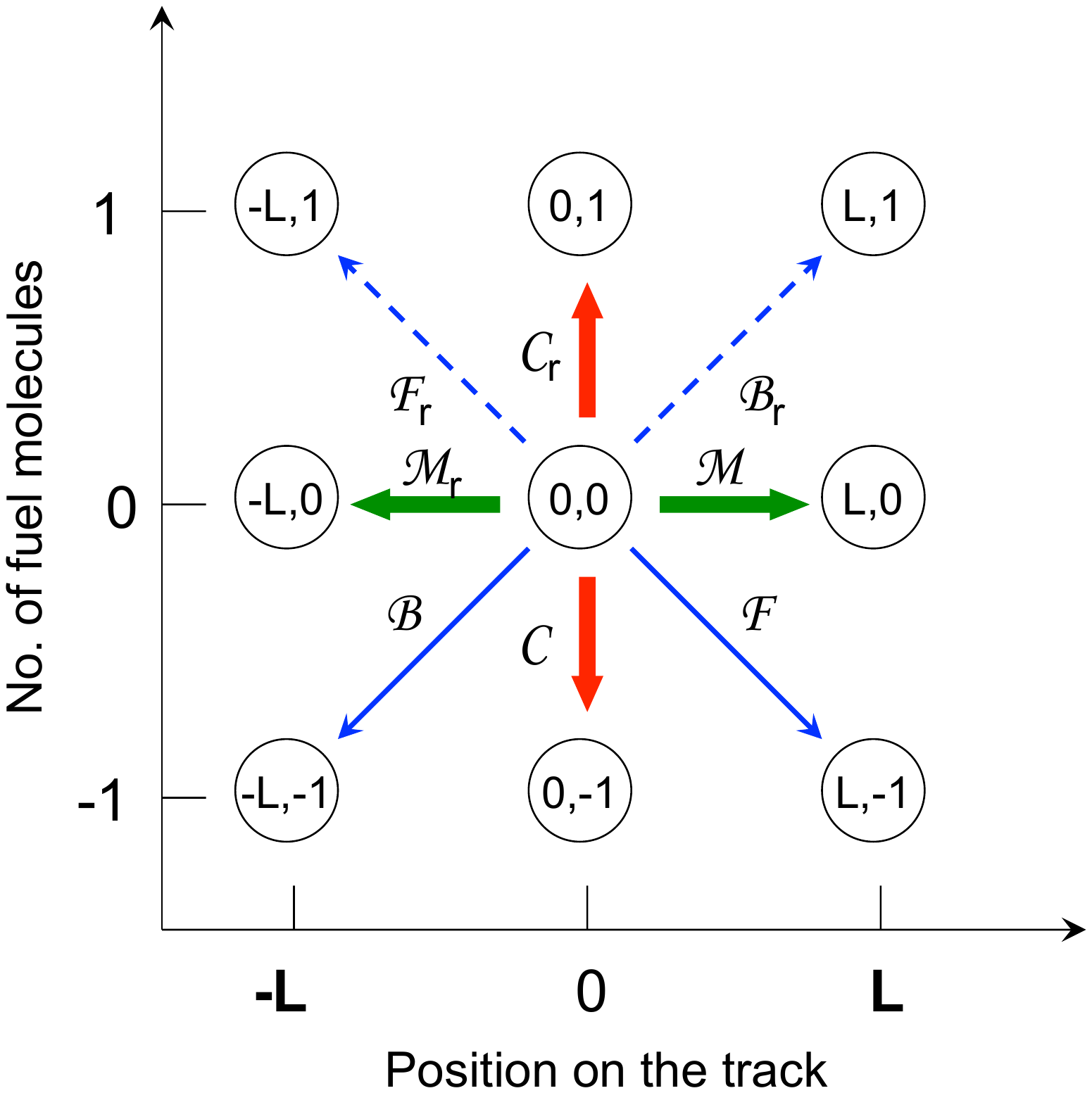}
\end{center}
\caption{Possible changes in the position and the number of ATP molecules in a cycle (adapted from ref.\cite{chowdhury13a}; see text for details).
}
\label{fig-mechchem}
\end{figure}

In each cycle, the position of the motor can change by $\pm L$ where $L$ is the fixed step size of the motor. Starting from any arbitrary initial condition with given position of the motor and given number of ATP molecules, all the eight possible transitions can be depicted schematically as shown in Fig.\ref{fig-mechchem}. The passive transition ${\cal M}$ and the corresponding reverse transition ${\cal M}_r$ are purely mechanical processes which do not change the number of ATP molecules. Similarly, the transitions ${\cal C}$ and ${\cal C}_r$ are purely chemical processes which do not change the position of the motor. But, both the active forward step (${\cal F}$) or the active backward step (${\cal B}$) are driven by hydrolyzing one molecule of ATP (see fig.\ref{fig-mechchem}). Does this motor synthesize, rather than hydrolyze, a molecule of ATP in the corresponding ``reverse processes'' (${\cal F}_r$ and ${\cal B}_r$), respectively \cite{chowdhury13a} (see fig.\ref{fig-mechchem})? Since addressing such fundamental questions on thermodynamics and kinetics of molecular motors \cite{block07} is beyond the scope of this article, interested readers are referred to the discussion in ref.\cite{astumian10}.

\begin{figure}[htbp]
\begin{center}
\includegraphics[width=0.95\columnwidth]{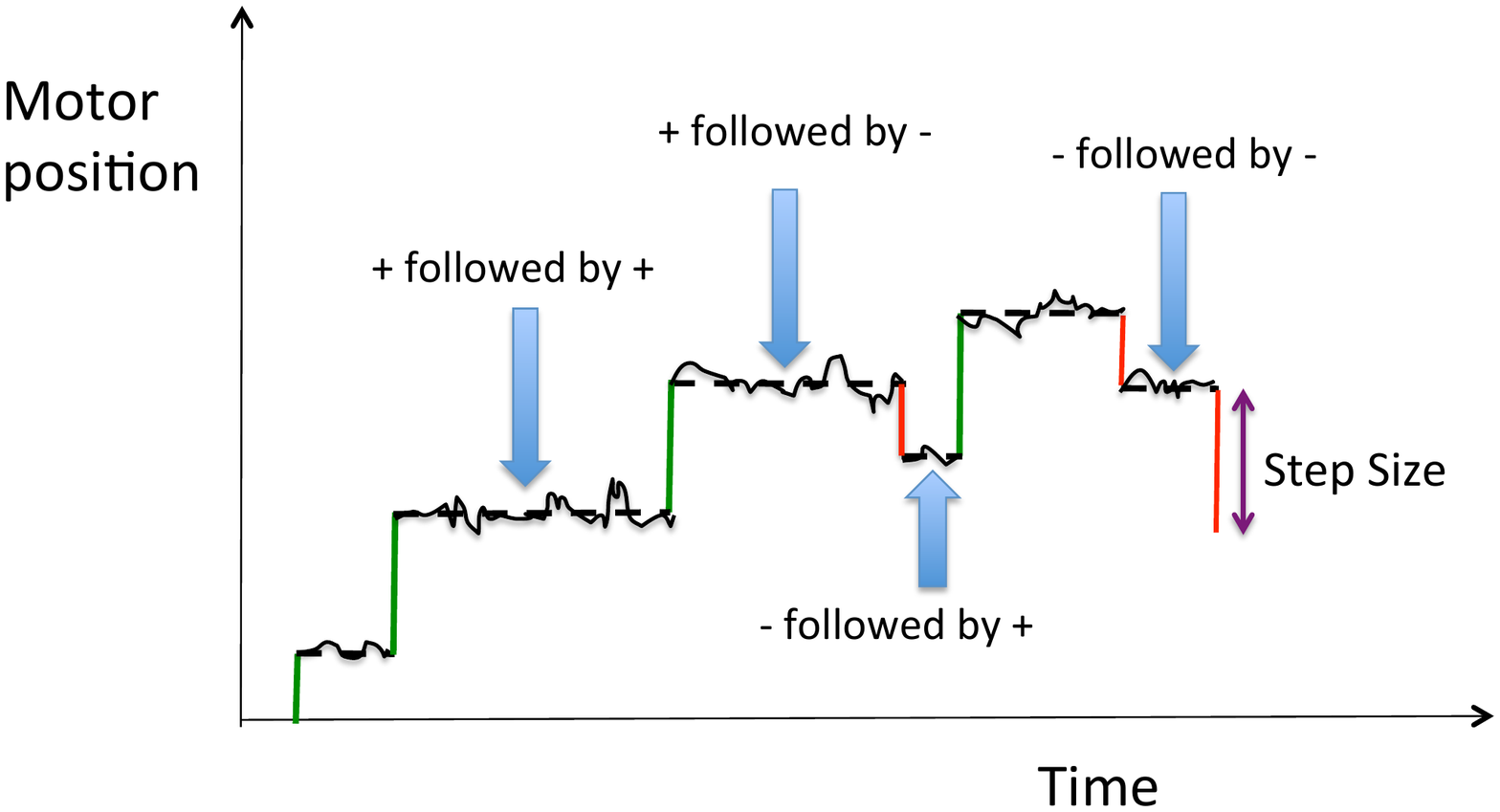}
\end{center}
\caption{A schematic depiction of the typical trace of the positions of a molecular motor with the progress of time. Both forward (+) and backward (-) stepping take place although the latter are quite rare.
}
\label{fig-conddwell}
\end{figure}

Thus, the stepping pattern of a molecular motor can be more complex than the progress of an enzymatic reaction. For example, a motor that can step both forward and backward, a forward step may coincide with the formation of the product of an enzymatic reaction catayzed by it whereas its backward step may coincide with the reformation of the substrate \cite{moffitt10b}. Thus, the complete mechano-chemical state attained by a forward step, for example, depends on the direction in which the preceding step was taken. For such motors dwell times corresponding to different kinetic pathways need to be sorted out for appropriate analysis of their distributions. The stochastic stepping of a motor that can step both forward and backward is characterised most appropriately in terms of the distributions of the four conditional dwell times
$\tau_{\pm \pm}$ which are defined as follows (see Fig.\ref{fig-conddwell}):
\begin{eqnarray}
\tau_{++} &=& {\rm dwell ~time ~between ~a ~+ ~step ~followed ~by ~a ~+ ~step} \nonumber \\
\tau_{+-} &=& {\rm dwell ~time ~between ~a ~+ ~step ~followed ~by ~a ~- ~step} \nonumber \\
\tau_{-+} &=& {\rm dwell ~time ~between ~a ~- ~step ~followed ~by ~a ~+ ~step} \nonumber \\
\tau_{--} &=& {\rm dwell ~time ~between ~a ~- ~step ~followed ~by ~a ~- ~step} \nonumber \\
\end{eqnarray}
where $+$ and $-$ refer to the forward and backward steps, respectively.
Moreover, at least for some motors, step size is not a constant; the distribution of step sizes itself is an interesting characteristic of the motor. Furthermore, because of the unavoidable noise in the recordings, extracting the dwell time requires careful processing of the data. However, new ideas are needed to gainfully exploit these conditional dwell times to extract useful information on the architecture of the network of the mechano-chemical states of the motor and kinetic pathways on that network.

%%%%%%%%%%%%%%%%%%%%%%%%%%%%%%%%%%%%%%%%%%%%%%%%%%%%%%%%%%%%%%%%%%%
\section{Allosteric mechanisms of intra-machine communication and the transmission system }
%%%%%%%%%%%%%%%%%%%%%%%%%%%%%%%%%%%%%%%%%%%%%%%%%%%%%%%%%%%%%%%%%%%

Allostery is a mechanism for regulation of the structure, dynamics and function of an enzyme by the binding of another molecule, called effector, which can be a small molecule (a ligand) or another macromolecule \cite{monod65,koshland66,changeux12,koshland02}.
The three defining characteristics of allostery are \cite{fenton08}: (i) the effector is chemically distinct from the substrate, (ii) the binding site for the effector is spatially well separated from that of the substrate, and (iii) binding of an effector molecule affects at least one of the functional properties of the enzyme; the functional property could be either (a) the binding affinity for its specific substrate or (b) the rate of the reaction it catalyzes.

The affinity of the motor for its track varies depending on the absence or presence (and nature) of the ligand (that is, ATP or ADP, etc.) in the ``engine'' of the motor.  The ATPase cycle of the engine, and the associated cyclic conformational change, is coordinated with the cyclic alteration of the motor's track-binding affinity. In general, on a given motor, the ``engine'' (where ATP gets hydrolyzed) and the site that binds to the track do not overlap and communicate with each other by allosteric mechanism \cite{goldsmith96,vologodskii06}.  
The hydrolysis of ATP causes only sub-nanometer movements in the ATPase engine. On the other hand, the step size of the cytoskeletal motors are at least an order of magnitude longer. The designs of the cytoskeletal motors are such that a ``mechanical element'', that extends from the engine, amplifies the subnanometer movements in the ATPase engine site up to step size of the motor \cite{vale00,llinas12}. This mechanism is analogous to the transmission system of an automobile. The component devices of the transmission system include, for example, the piston, shaft, clutch and gears, etc. and it transmits the engine-generated power to the wheels. 

Both myosins and kinesins use a ``sensor'' to detect the presence or absence of a single phosphate group.
The sensor consists of loops called switch I and switch II. To serve as the specific sensor for the phosphate group, the switch II loop swings in and out, respectively,in the presence and absence of the phosphate group. When ATP is hydrolyzed to ADP  the small movements of the sensor attached to the engine of the motor is transmitted to the track-binding site by a helix that is quite long and rigid. In both kinesins and myosins the conformational changes of the helix have, at least, a superficial similarities with the movement of a piston. The inward motion of the switch II towards the phosphate causes the upstroke of the piston whereas phosphate release initiates the downstroke. The mechanical elements of kinesin and myosin, called neck-linker and lever arm, respectively, executes movements driven by the corresponding piston-like helices described above. In myosin, the motion of the helix drives the angular motion of the lever arm that, in turn, gives a push to an actin filament . In kinesin movement of the the piston-like helix causes docking of the neck linker which, in turn, is responsible for the coordinated movement of the two heads of the kinesin motor in a directed manner along a MT track.

Investigations on the pathways (amino acid sequence) for intra-machine allosteric communication began only in the recent past. Although normal mode analysis of  coarse-grained theoretical models of the motors \cite{zheng09} is a very popular theoretical technique for identifying the allosteric communication pathways, other alternative methods have also been used (see, for example, ref.\cite{tang07} for treatment of myosin)

Establishing the mechanism of communication between the engine and track-binding site of a dynein is more challenging than that for most of the other molecular motors. Out of the six domains that form a ring-like structure, only two seem to operate as engines by hydrolysing ATP. But, the track-binding site is not located  anywhere on this ring. Instead, the track-binding site is located at the end of a 15 nm long stalk that emerges from this ring. One of the biggest mysteries in the designs of cytoskeletal motors is how the sub-nanometer conformational change in the engine of dynein is allosterically communicated to the tip of the stalk \cite{spudich11,cho12}. Experimental data for dynein have been interpreted to suggest that it also has gears \cite{mallik04a} and clutch \cite{huang12}; allosteric interactions are believed to be responsible for the gear and clutch mechanisms of dynein.

%%%%%%%%%%%%%%%%%%%%%%%%%%%%%%%%%%%%%%%%%%%%%%%%%%%%%%
\section{Summary and conclusion}
%%%%%%%%%%%%%%%%%%%%%%%%%%%%%%%%%%%%%%%%%%%%%%%%%%%%%%

In this commemorative article we have discussed some topics of current research interest in molecular motors from the perspective of biochemistry. In particular, we have examined the roles of Michaelis-Menten scheme of enzymatic reactions and allosteric mechanism of intra-molecular communication, respectively, in the operation of the engine and transmission system of the cytoskeletal motors. 
A class of motors, that includes ribosome, use their nucleic acid tracks also as a template for polymerizing other macromolecules. Replication of DNA, transcription and translation are examples of such template-directed polymerisation. The machines that drive such phenomena have the daunting task of catalyzing the polymerization reaction with much higher fidelity than what might be demanded by the laws of thermodynamics. The amplification of substrate specificity of these enzymes is achieved through kinetic processes like kinetic proofreading, energy relay, etc. How these machines optimise the opposing demands for speed and fidelity is one of the interesting questions at the interface of enzymatic kinetics and biophysics of molecular motors \cite{sharma10}.

The open questions on the mechanism of operation of the transmission system of a motor are closely related to the fundamental general questions on the physical modes of long-distance communication in macromolecular systems \cite{nussinov12}. For example, does nature select the path of shortest distance or shortest time for allosteric communication? Better methods of characterization of the pathways of allosteric interactions may reveal generic principles that nature might have exploited in designing the transmission systems through evolutionary tinkering. 

Although equipped with a noisy engine and a shaky transmission system, a molecular motor moves, on the average, in a directed manner in a molecular hailstorm- a remarkable feat, indeed. Explaining the physical origin of the MM-like expression for its average velocity and understanding the allosteric mechanism of the transmission system are some important items in the overall agenda of research on molecular motors \cite{chowdhury13a,chowdhury13b,pollard03a}. Michaelis-Menten at 100 is as exciting, if not more, as it was in 1913. The concept of allostery at 50 is, perhaps, finding applications in more diverse systems than might have been anticipated in 1963. Happy birthday allostery! Long live Michaelis-Menten!

%%%%%%%%%%%%%%%%%%%%%%%%%%%%%%%%%%%%%%%%%%%%%%%%%%%%%%
\noindent{\bf Acknowledgements:}
%%%%%%%%%%%%%%%%%%%%%%%%%%%%%%%%%%%%%%%%%%%%%%%%%%%%%%

I thank Ajeet K. Sharma for a critical reading of the manuscript.
This work is supported by Dr. Jag Mohan Chair professorship, by J.C. Bose national fellowship, and by a research grant from DBT, government of India. I also thank the visitors program of the Max-Planck Institute for the Physics of Complex Systems for hospitality in Dresden where part of this article was written.

%%%%%%%%%%%%%%%%%%%%%%%%%%%%%%%%%%%%%%%%

\clearpage

%%%%%%%%%%%%%%%%%%%%%%%%%%%%%%%%%%%%%%%%%%%%%%%%%%%%%%%%%%%%%%
%%%%%%%%%%%%%%%%%%%%%%%%%%%%%%%%%%%%%%%%%%%%%%%%%%%%%%%%%%%%%

%%%%%%%%%%%%%%%%%%%%%%%%%%%%%%%%%%%%%%%%%%%%%%%%%%%%%%%%%%%%%%
%%%%%%%%%%%%%%%%%%%%%%%%%%%%%%%%%%%%%%%%%%%%%%%%%%%%%%%%%%%%%

%%%%%%%%%%%%%%%%%%%%%%%%%%%%%%%%%%%%%%%%%%%%%%%%%%%%%%%%%%%%%%
%%%%%%%%%%%%%%%%%%%%%%%%%%%%%%%%%%%%%%%%%%%%%%%%%%%%%%%%%%%%%

%%%%%%%%%%%%%%%%%%%%%%%%%%%%%%%%%%%%%%%%%%%%%%%%%%%%%%%%%%%%%%
%%%%%%%%%%%%%%%%%%%%%%%%%%%%%%%%%%%%%%%%%%%%%%%%%%%%%%%%%%%%%

%%%%%%%%%%%%%%%%%%%%%%%%%%%%%%%%%%%%%%%%%%%%%%%%%%%%%%%%%%%%%%
%%%%%%%%%%%%%%%%%%%%%%%%%%%%%%%%%%%%%%%%%%%%%%%%%%%%%%%%%%%%%

%%%%%%%%%%%%%%%%%%%%%%%%%%%%%%%%%%%%%%%%%%%%%%%%%%%%%%%%%%%%%%
%%%%%%%%%%%%%%%%%%%%%%%%%%%%%%%%%%%%%%%%%%%%%%%%%%%%%%%%%%%%%

%%%%%%%%%%%%%%%%%%%%%%%%%%%%%%%%%%%%%%%%%%%%%%%%%%%%%%%%%%%%%%
%%%%%%%%%%%%%%%%%%%%%%%%%%%%%%%%%%%%%%%%%%%%%%%%%%%%%%%%%%%%%

\clearpage 

%%%%%%%%%%%%%%%%%%%%%%%%%%%%%%%%%%%%%%%%%%%%%%%%%%%%%%%%%%%%%%%%%%%%%
\begin{table}
\begin{tabular}{|c|c|c|c|} \hline
Motor family & Track & Example of function & Step size \\ \hline \hline
Myosin-V \& VI & F-actin & Melanosome transport & 36 nm \\ \hline
Kinesin-1 \& cytoplasmic dynein & Microtubule & Mitochondria transport & 8 nm \\ \hline
\end{tabular}
\caption{Few examples of ATP-driven porters are listed along with their
corresponding tracks. The actual step size can be an integral multiple
of the minimum value quoted in this table (see ref.\cite{chowdhury13a} for references to original papers).
 }
\label{table-porters}
\end{table}
%%%%%%%%%%%%%%%%%%%%%%%%%%%%%%%%%%%%%%%%%%%%%%%%%%%%%%%%%%%

%%%%%%%%%%%%%%%%%%%%%%%%%%%%%%%%%%%%%%%%%%%%%%%%%%%%%%%%%%%%%%%%%%%%%
%%%%%%%%%%%%%%%%%%%%%%%%%%%%%%%%%%%%%%%%%%%%%%%%%%%%%%%%%%%%%%%%%%%%%
\begin{table}
\begin{tabular}{|c|c|c|} \hline
Motor family  & Filament & Example of function \\\hline \hline
Myosin-II  & F-actin & Muscle contraction \\ \hline
Kinesin-5, Kinesin-14  & MT & Mitosis \\ \hline
Axonemal Dynein  & MT & Flagellar beating \\ \hline
\end{tabular}
\caption{Few example of rowers and sliders (see ref.\cite{chowdhury13a} for references to original papers).
 }
\label{table-rowslide}
\end{table}
%%%%%%%%%%%%%%%%%%%%%%%%%%%%%%%%%%%%%%%%%%%%%%%%%%%%%%%%%%%

%%%%%%%%%%%%%%%%%%%%%%%%%%%%%%%%%%%%%%%%%%%%%%%%%%%%%%%%%%%%%%%%%%%%%
\begin{table}
\begin{tabular}{|c|c|c|c|c|} \hline
Motor  & Template & Product & Function & Step-size \\ \hline \hline
DdDP  & DNA & DNA & DNA replication & 0.34 nm (1 nt) \\ \hline
DdRP  & DNA & RNA & Transcription & 0.34 nm (1 nt) \\ \hline
RdDP  & RNA & DNA & Reverse transcription & 0.34 nm (1 nt) \\ \hline
RdRP  & RNA & RNA & RNA replication & 0.34 nm (1 nt) \\ \hline
Ribosome  & mRNA & Protein & Translation & 1.02 nm (3 nt) \\ \hline
\end{tabular}
\caption{Types of polymerizing machines, the templates that direct 
the sequence of monomers of the polymeric products. The abbreviations DdDP, DdRP, RdDP and RdRP refer to DNA-dependent DNA polymerase, DNA-dependent RNA polymerase, RNA-dependent DNA polymerase and RNA-dependent RNA polymerase, respectively (see ref.\cite{chowdhury13a} for references to original papers).
 }
\label{table-polymerase}
\end{table}
%%%%%%%%%%%%%%%%%%%%%%%%%%%%%%%%%%%%%%%%%%%%%%%%%%%%%%%%%%%

\end{document}